\documentclass[aps,prb,twocolumn,showpacs]{revtex4}
\usepackage{graphicx}

\begin{document}

\title{Quantum dot nucleation in strained-layer epitaxy: minimum-energy
pathway in the stress-driven 2D-3D transformation}

\author{Jos\'e Emilio Prieto}
\email{joseemilio.prieto@uam.es}
\affiliation{Centro de Microan\'alisis de Materiales and Instituto
Universitario ``Nicol\'as Cabrera'', Universidad Aut\'onoma de Madrid, E-28049
Madrid, Spain}

\author{Ivan Markov}
\email{imarkov@ipchp.ipc.bas.bg}
\affiliation{Institute of Physical Chemistry, Bulgarian Academy of Sciences,
1113 Sofia, Bulgaria}
\date{\today}

\begin{abstract}
The transformation of monolayer islands into bilayer islands as a first 
step of the overall two-dimensional to three-dimensional (2D-3D) 
transformation in the coherent Stranski-Krastanov 
mode of growth is studied for the cases of expanded and compressed overlayers.
Compressed overlayers display a nucleation-like behavior: the
energy accompanying the transformation process displays a maximum at some
critical number of atoms, which is small for large enough values of the 
misfit, and then decreases gradually down to the completion of the 
transformation, non-monotonically due to the atomistics of the process. 
On the contrary, the energy change in expanded overlayers 
increases up to close to the completion of the transformation 
and then abruptly collapses with the disappearance of the monoatomic 
steps to produce low-energy facets. This kind of transformation takes place
only in materials with strong interatomic bonding. Softer materials under
tensile stress are expected to grow predominantly with a planar morphology
until misfit dislocations are introduced, or to transform into 3D islands
by a different mechanism. It is concluded that the coherent Stranski-Krastanov 
growth in expanded overlayers is much less probable than in compressed 
ones for kinetic reasons.

\end{abstract}


\maketitle

\section{INTRODUCTION}

Understanding the mechanism of transition from a pla\-nar two-dimensional
(2D) thin film to a three-di\-men\-si\-o\-nal (3D) island morphology in the
he\-te\-ro\-epi\-taxy of highly mismatched materials is of crucial importance
for the growth of self-assembled quantum dots in nanoscale technology. In
the {\em coherent} Stranski-Krastanov (SK) mode of growth, dislocation-free 
3D islands develop on top of a 2D wetting layer in order to relieve
the misfit strain at the expense of an increase in surface energy.\cite{Eag} 
This mechanism of strain
relaxation is established in a multitude of systems of technological
importance for the manufacturing of optoelectronic devices.\cite{Bruce} Despite
the huge amount of studies devoted to the evolution of the cluster shape,
many aspects, in par\-ti\-cu\-lar the very beginning of the 2D-3D 
transition, still remain unclear.

The first theoretical concept for the transition from a 2D layer to faceted
3D islands included a nucleation mechanism as a result of the interplay
between the surface energy and the relaxation of the strain energy relative
to the values of the wetting layer.\cite{Jerr} Irreversible 3D growth 
was predicted 
to begin above a critical volume, overcoming an energetic barrier whose
height is inversely proportional to the forth power of the lattice misfit.

Mo {\it et al.} observed Ge islands representing elongated pyramids (``huts'')
bounded by \{105\} facets inclined by 11.3$^{\circ}$ to the substrate.\cite{Mo}
These clusters were thought to be a step in the pathway to the formation of
larger islands with steeper side walls (``domes'' or ``barns'').\cite{Sut} Chen
{\it et al.} studied the earliest stages of Ge islanding and found that
Ge islands smaller than the hut clusters do not involve discrete \{105\}
facets.\cite{Chen} This result was later confirmed by Vailionis {\it et al.}
who observed the formation of 3-4 monolayers-high ``prepyramids'' with rounded
bases which existed over a narrow range of Ge coverages in the beginning of 
the 2D-3D transformation.\cite{Vai} Sutter and Lagally\cite{Sut2} assumed
that faceted, low-misfit alloy 3D islands can result from morphological
instabilities (ripples) that are inherent to strained films,\cite{AT,G,S} thus
suggesting that 3D islands can be formed without the necessity to overcome a
nucleation barrier. Similar views were simultaneously expressed by Tromp {\it
et al.}\cite{Tromp} Tersoff {\it et al.} developed further this idea
suggesting that the transition from the initial smooth ``prepyramids'' to
faceted pyramids can be explained by assuming that the polar diagram of SiGe
alloy islands allows the existence of all orientations vicinal to (001) with
the first facet being \{105\}.\cite{Jerry1}

In order to explain the experimental observation that SiGe alloy films roughen
only under compressive stresses larger than a critical value of 1.4~\%, Xie
{\it et al.} assumed that the smallest 3D islands have stepped rather than
faceted side surfaces.\cite{Xie1} They noted that the steps on the SiGe(001)
vicinals are under tensile stress and the\-ir en\-er\-gy of formation is
lowered by the
compressive misfit but increased by the tensile strain [see also the
discussion in Refs. (\onlinecite{Jerry2}) and (\onlinecite{Xie2})]. As a
result, the step formation and in turn the roughening are favored by the
compressive misfit. It is worth noting
that a barrierless evolution of stepped islands was predicted by Sutter and
Lagally under the assumption that the slope of the side walls of the stepped
islands increases continuously from zero to 11.3$^{\circ}$.\cite{Sut2}

Priester and Lannoo suggested that 2D islands of monolayer height appear 
as precursors of the 3D is\-lands.\cite{Pri} In addition, it was 
established that the mi\-ni\-mum-energy pathway of the 2D-3D transition 
has to consist of a series 
of intermediate states with thicknesses increasing in monolayer steps
and which are stable in separate intervals of volume. The first step in this
transformation should be the rearrangement of mo\-no\-lay\-er into bilayer
islands.\cite{Kor,Prieto} Khor and Das Sarma found by Monte Carlo simulations
that during the re\-ar\-range\-ment, the material for the bilayer island comes
almost completely from the original monolayer island, the bulk of the
material for the three-layer island comes from the original two-layer island,
etc.\cite{Khor}

Moison {\it et al.} reported that the coverage suddenly decreases from about 
1.75 ML to 1.2 ML when 3D InAs islands begin to form on GaAs.\cite{Moi} 
The same phenomenon has been observed by Shklyaev {\it et al.} in the case of
Ge/Si(111).\cite{Shkl} These observations suggest a process of rearrangement
as mentioned above. Voigtl\"ander and Zinner noted that 3D Ge islands
have been observed in the same locations on the Si(111) surface where 2D
islands locally exceeded the critical thickness of the wetting layer of two
bilayers.\cite{Voi} One-monolayer thick InAs islands were suggested to act
as precursors for formation of thicker structures on GaAs.\cite{Polimeni}
The simultaneous pre\-sence of stable one, two, three or four
monolayers-thick islands has been observed in heteroepitaxy of InAs on InP
and GaAs.\cite{Rud1,Rud2,Col}

In this paper we studied the earliest stages of growth of thin films
in the coherent (dislocation-free) Stranski-Krastanov mode. We considered
the in\-sta\-bility of the planar growth against clustering
by focussing on the conservative (i.e., without considering 
further de\-po\-si\-tion) mono- to bilayer transformation as a first step 
of the overall 2D-3D transition, or the beginning of the formation
of the ``prepyramids'' mentioned above. We found that this transformation
is a true nuc\-le\-a\-tion process in compressed overlayers, in the sense 
that a critical nucleus of the second layer is initially formed and then
grows further up to the complete mono-bilayer transformation. The energy 
associated with the transformation thus reaches a maximum and then  
starts a decreasing trend.\cite{CGB} This is not the case in 
expanded overlayers, where 
the energy tends to increase up to very close to the completion of the 
transformation and then steeply decreases at the very end. 
The main result of this study is that coherent Stranski-Krastanov
growth in expanded overlayers is much less probable than in compressed ones.

\section{Model}

We consider an atomistic model in $2+1$ dimensions. The 3D crystallites have
{\em fcc} structure and (100) surface orientation, thus possessing the 
shape of truncated squ\-are py\-ra\-mids. 
We found that monolayer-high elon\-gat\-ed islands always have higher energy 
than square islands with the same number of atoms, as expected from the 
symmetry of the lattice and the isotropy of the interaction potentials
(see below).  The lattice misfit is the same in both orthogonal di\-rec\-tions. 
We consider interactions only in the first coordination sphere; inclusion of
further coordination spheres does not alter qualitatively the results.

The choice of crystal lattice and interaction potential is more 
appropriate for the heteroepitaxy of (close-packed) metals on metals
rather than for that of se\-mi\-con\-duc\-tor materials. 
As a consequence, properties that depend crucially on the strong
directional bonding characteristic of semiconductors cannot be addressed
by our model. Some examples are: the dependence of the shape of GeSi/Si
dots on volume~\cite{Mo,Sut,Chen} as discussed above; the observation 
of lens-shaped~\cite{Don,Wal} and pyramidal~\cite{Heyn} dots and even 
of coexistence of both types~\cite{Bhatti} in InAs/GaAs, the other 
well-studied system (for a recent review see Ref.~\onlinecite{Bruce2}) 
or the cases where the accommodation of the lattice misfit of a given 
material on different crystallographic faces of the same substrate takes 
place by other mechanisms (also found for InAs/GaAs),~\cite{Joyce} where 
additional aspects as the presence of different surface reconstructions 
affect the thermodynamical balance of surface energies as well as the 
diffusion kinetics and, as a consequence, the nucleation behaviour 
and the growth mode.
As our aim is to study the ``reversible" minimum-energy pathway of the 
transition from metastable states to the ground state of a
given system, the exact particularities of the model are not likely to play 
a crucial role and we expect the same qualitative behavior for any crystal 
lattice, crystal shape and interatomic potential.

We have performed atomistic calculations making use of a simple minimization
procedure. The atoms interact through a pair potential whose
an\-har\-mo\-ni\-ci\-ty can be varied by adjusting two constants $\mu $ and
$\nu $ ($\mu  > \nu $) that govern separately the repulsive and the
attractive branches, respectively,\cite{Mar1,Mar2}
\begin{eqnarray}\label{potent}
V(r) = V_{o}\Biggl[\frac{\nu }{\mu - \nu }e^{-\mu (r-b)} - \frac{\mu }{\mu -
\nu }e^{-\nu (r-b)}\Biggr],
\end{eqnarray}
where $b$ is the equilibrium atom separation. For $\mu  = 2\nu $ the potential
(\ref{potent}) turns into the familiar Morse form. 

A static relaxation
of the system is performed by allowing each atom to displace in the direction
of the force, i.e., the gradient of the energy with respect to the atomic 
coordinates, in an iterative procedure until all the forces fall below some
negligible cutoff value. As we were only interested in the 2D-3D
transformation of isolated islands, the calculations were performed
under the assumption that the substrate (the wetting layer) is rigid;
the atoms there are separated by a distance $a$. The lattice misfit is
thus given by $\varepsilon=(b-a)/a$.

\begin{figure}[htb]
\includegraphics*[width=7.5cm]{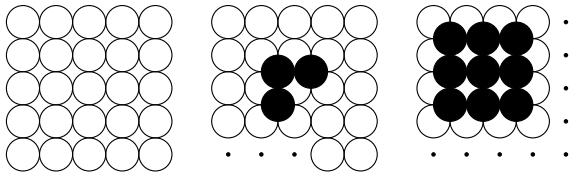}
\caption{\label{mech} Schematic process for the evaluation of the
activation energy of the monolayer-bilayer transformation.
The initial state is a square monolayer island. The intermediate state is a
monolayer island short of some number of atoms which are detached from the
edges and placed in the second level. The final state is a truncated bilayer
pyramid.}
\end{figure}

For the study of the mechanism of mono-bilayer trans\-for\-ma\-tion, we assume 
the following imaginary model process:\cite{Stmar} atoms detach from the 
edges of
monolayer islands, which are larger than the critical size for the mono-
to bilayer transformation $N_{12}$ and thus unstable against
bilayer islands, diffuse on top of them, aggregate and give rise
to second-layer nuclei. These grow at the expense of the atoms detached
from the edges of the lower islands. The process continues up to the moment
when the upper island completely covers the lower-level island. To simulate
this process, we assume an initial square monolayer island, detach atoms
one by one
from its edges and locate them on top and at the center of the ML island,
building there structures as compact as possible (Fig. \ref{mech}).
The energy change associated with the process of transformation at a 
particular stage is given by the difference between the energy of the 
incomplete bilayer island and that of the 
initial monolayer island. This is in fact a conservative version of the 
mechanism observed by Khor and Das Sarma in 1+1 dimensions.\cite{Khor}

\section{Results}

\subsection{Stability of Monolayer Islands}

In our previous work in (1+1)D models, it was es\-tab\-lish\-ed that
monolayer-high islands are stable against bilayer islands up to 
some critical volume or number of atoms $N_{12}$; in turn, bilayer 
islands are stable against trilayer islands up to
another critical number $N_{23} > N_{12}$; etc.\cite{Kor,Prieto} The
mono-bilayer transformation was considered as the first step of the overall
2D-3D transformation and a critical misfit was determined from the misfit
dependence of $N_{12}$. The latter was found to increase with decreasing
misfit diverging at a critical value $\varepsilon _{12}$. Above the critical
misfit, the coherent Stranski-Krastanov mode is favored against the 
layer-by-layer growth followed by introduction of misfit dislocations.
The opposite is true below the critical value.
Whereas this critical behavior is clearly pro\-noun\-ced with com\-pressive
stra\-in, it is much smoother in expanded over\-lay\-ers. It is worth
noting that the existence of cri\-ti\-cal misfit was observed in a series of
heteroepitaxial systems.\cite{Xie1,Leo,Wal,Pinc}

\begin{figure}[htb]
\includegraphics*[width=8.5cm]{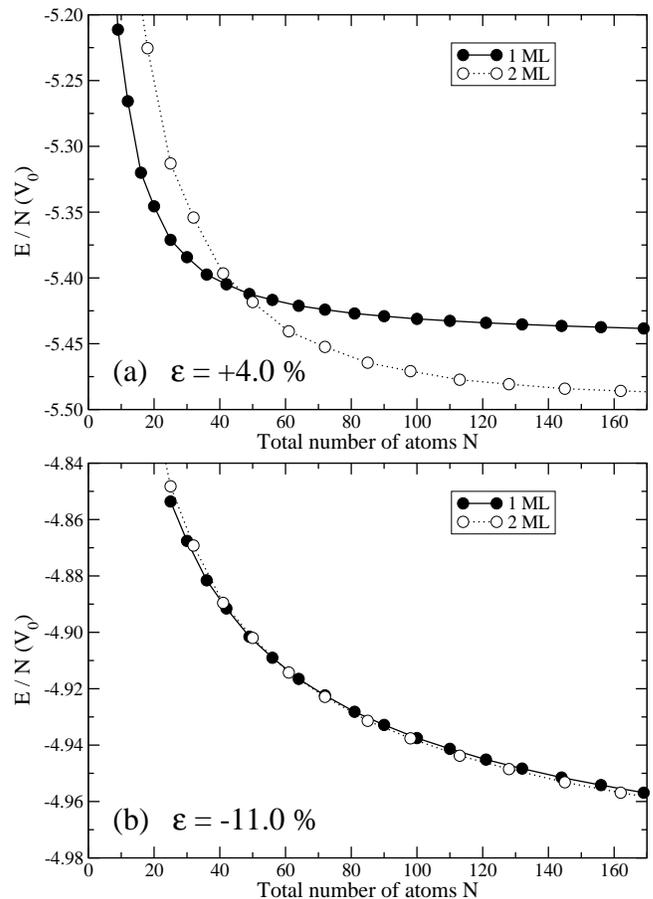}
\caption{\label{e12} Total energy per atom of mono- and bilayer islands
at (a) positive
(+4.0~\%) and (b) negative (-11.0~\%) values of the misfit as a function
of the total number of atoms. The potential is of the form given by
eq.~(\ref{potent}) with $\mu=16$ and $\nu=14$.}
\end{figure}

In the present work, using more realistic (2+1)D models, we found a larger
dif\-fer\-ence in the behavior of expanded and compressed overlayers. 
Figure \ref{e12} shows that the total energies of mono- and bilayer islands
under tensile stress containing the same total number of atoms are very
close to each
other compared with the corresponding behavior in compressed overlayers. As
will be shown below this leads to the conclusion that even for $N \gg N_{12}$,
determined as the crossing point of the curves corresponding to monolayer
and bilayer islands in Fig.~\ref{e12}, 
the probability of the 2D-3D trans\-formation remains nearly equal to the
probability of the reverse 3D-2D transformation.

\begin{figure}[htb]
\includegraphics*[width=8.5cm]{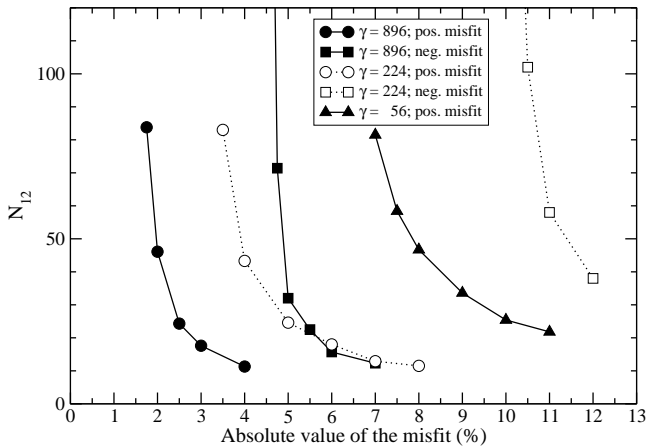}
\caption{\label{gamma} Critical island size $N_{12}$ (number of atoms)
as a function of the lattice mismatch at different values of the force
constant $\gamma = \mu \nu V_0$. Potentials of the form given
by eq.~(\ref{potent}) were used, with $\mu / \nu$ = 8/7 ($V_0$ is taken
equal to unity). As seen, coherent 3D islanding is favored in
expanded overlayers only in ``stiffer'' materials.}
\end{figure}

It turns out that the misfit dependence of $N_{12}$ is 
very sensitive to the value of the force constant 
$\gamma  = \mu \nu V_0$
of the interatomic bonds, particularly in expanded overlayers
(Fig. \ref{gamma}). Decreasing $\mu$ and $\nu$
($V_0$ is assumed equal to unity) in such a way that the ratio $\mu / \nu$
is kept constant (in this case equal to 8/7), shifts the intersection 
points $N_{12}$ to larger absolute values of the misfit. As a result, a 
critical size $N_{12}$ in compressed overlayers exists practically for all 
values of $\gamma $ whereas in overlayers under tensile stress, 
$N_{12}$ shifts to so large values of the misfit that they 
effectively disappear below some critical value of $\gamma $.

It was established that in the case of a force constant of an
intermediate value ($\mu = 2\nu = 12$), $N_{12}$ disappears (the monolayer
islands are always stable against bilayer islands), but $N_{13},
N_{14}, N_{23}...$ still exist. This points to a novel mechanism of 
2D-3D transformation which differs from the
consecutive formation of bilayer, trilayer, etc. islands, each from the
previous one. The new mechanism obviously consists of the formation and
2D growth of bilayer islands on top of the monolayer is\-land, thus 
transforming the initial monolayer island directly into a trilayer island.
At even smaller values of $\gamma$, the critical values $N_{13}, N_{14}$ etc. 
consecutively disappear, suggesting a generalized mechanism of 
2D-3D transformation 
in which the monolayer islands transform into thicker islands. 
This multilayer 2D mechanism will be a subject of a se\-par\-ate study. 
In this paper we focus on the layer-by-layer 2D-3D transformation.

\begin{figure}[htb]
\includegraphics*[width=8.5cm]{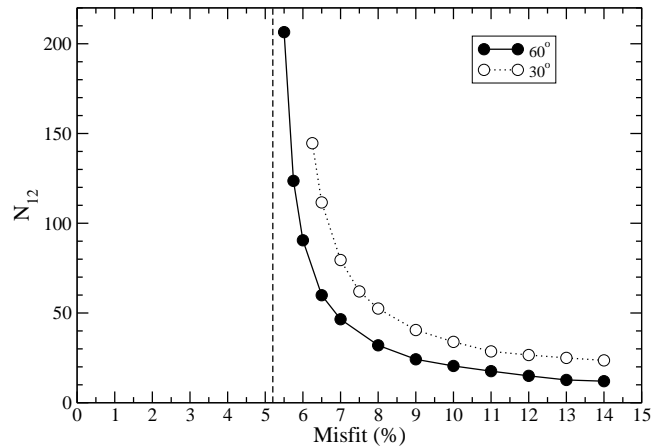}
\caption{\label{N12_2d} Misfit dependence of the critical sizes
$N_{12 }$ of mono- and bilayer islands with different shapes, given by
different angles of the side walls: 60$^{\circ}$ and 30$^{\circ}$
($\mu= 2 \nu = 12$). The critical misfit $\varepsilon_{12}$ is shown
by the vertical dashed line. }
\end{figure}

We considered also the stability of monolayer islands against bilayer islands
with a different slope of the side walls. It was found that $N_{12}$ is
smaller if the slope of the side walls is the steepest one (60$^{\circ}$) for
this lattice in comparison with flatter islands (Fig. \ref{N12_2d}). This is
due to the fact that in crystals with steeper side walls, the strain
relaxation is more efficient than in flatter islands. 
This is in contradiction with the experiments in semiconductor growth in which 
islands with side walls of a smaller slope than that of the first facet are
initially observed.~\cite{Chen,Vai,Xie1}
In any case this means that we can exclude the flatter islands from our 
consideration.

\begin{figure}[htb]
\includegraphics*[width=8.5cm]{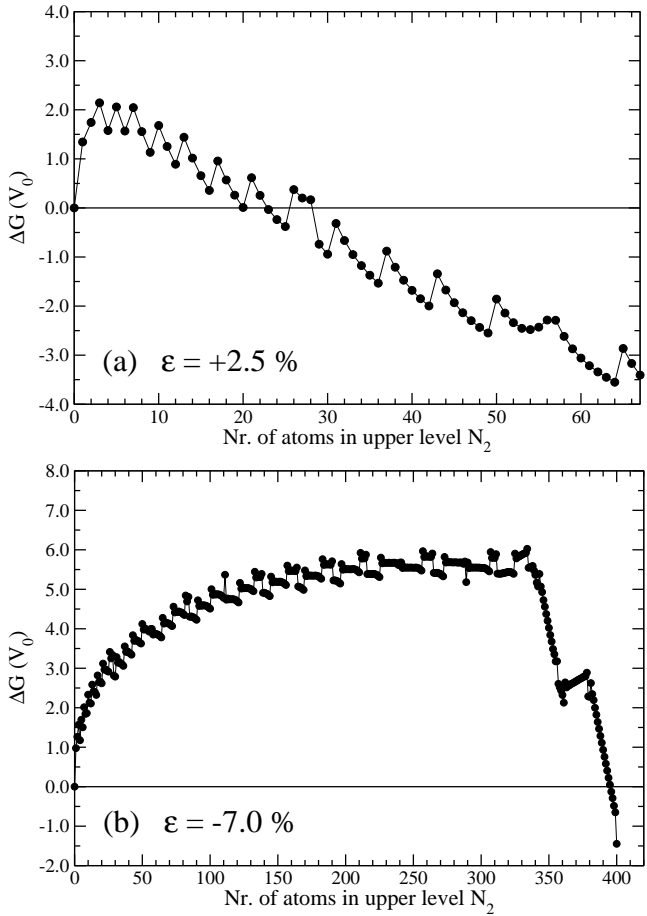}
\caption{\label{DeltaE} Transformation curves representing the energy change
in units of the bond energy $V_0$ as a function of the number of atoms in the
upper level for (a) positive (+2.5~\%) and (b) negative (-7.0~\%) values of the
misfit. The number of atoms in the initial monolayer island ($N_0 =
841 = 29\times29$) is chosen in such a way that the resulting truncated
bilayer pyramid is complete ($21\times21 = 441$ atoms in the lower and
$20\times20 = 400$ atoms in the upper level); $\mu = 2\nu = 36$.}
\end{figure}

We conclude that for some reasonable degree of an\-har\-mo\-ni\-ci\-ty 
(e.g. $\mu = 2\nu$ in our model), monolayer islands become unstable against 
bilayer islands thus making possible the 2D-3D transformation by the 
layer-by-layer mechanism only at strong enough interatomic bonding. 
Soft materials are ex\-pected to grow either
with a planar morphology until misfit dislocations are introduced or to
transform into 3D islands by a different, multilayer 2D mechanism.

\subsection{Mechanism of 2D-3D transformation}

Figure \ref{DeltaE} shows typical transformation curves of the energy change
as a function of the number of atoms in the upper island for (a) positive 
and (b) negative
misfits. It is immediately seen that in a compressed overlayer, 
the transformation curve for $\Delta G$ has the typical shape
of a nucleus formation curve: it displays a maximum $\Delta G_{max}$ for a
cluster consisting of a small number of atoms $n_{max}$ and then
decreases beyond this size up to the completion of the transformation.
The atomistics of the transfer process (i.e. the completion of rows in the
upper level and their depletion in the lower one) is responsible for the
jumps (the non-monotonic behaviour) in the curve. 

\begin{figure}[htb]
\includegraphics*[width=8.5cm]{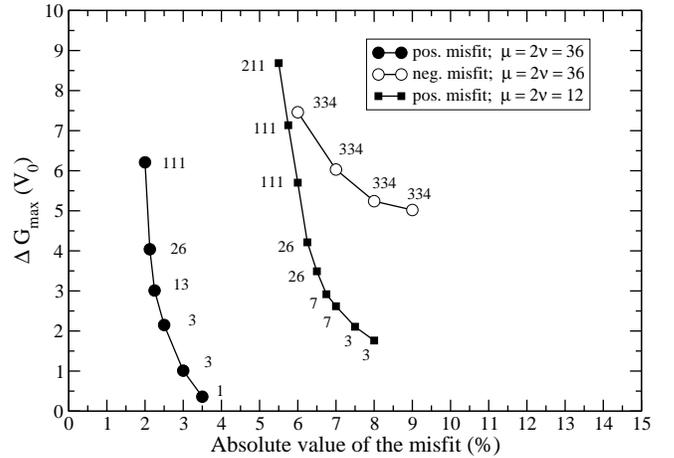}
\caption{\label{height} Height of the energetic barriers in units of $V_0$
as a function of the absolute value of the lattice misfit.
The figures at each point show the number of atoms in the critical nucleus
$n_{max}$. The initial island size was 29$\times$29 = 841 atoms.  The round
symbols were calculated for $\mu=2\nu=36$, the squares for $\mu=2\nu=12$.}
\end{figure}

In the case of expanded overlayers, the energy change increases up to
a large number of atoms (again non-monotonically due to the atomistics) 
and then abruptly decreases at the end of the
transformation. No true maximum is displayed. The energy change becomes
negative only after the transfer of the last several atoms. Comparing the
largest value of the energy with the energy at the transformation completion
leads to the conclusion that the probabilities of the direct and reverse
transformations are nearly equal.

\begin{figure}[htb]
\includegraphics*[width=8.5cm]{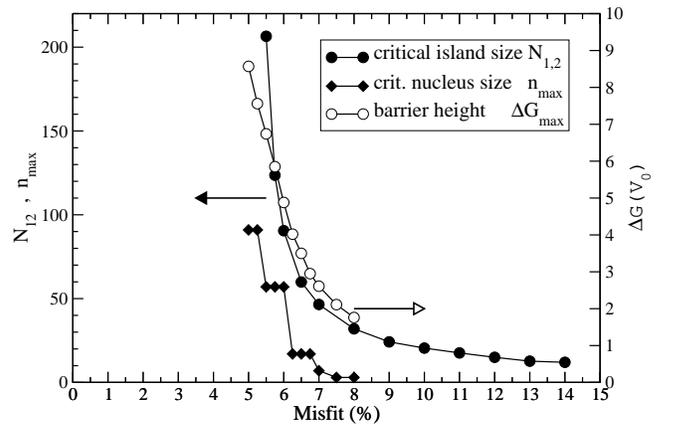}
\caption{\label{n12dg} Misfit dependence of the critical island size $N_{12}$,
the critical nucleus size $n_{max}$, (both expressed in number of atoms),
and the nucleation barrier height $\Delta G_{max}$ (in units of $V_0$)
for compressed overlayers and $\mu=2\nu=12$. The last two magnitudes were
computed for islands of an initial size of 20$\times$20 = 400 atoms.}
\end{figure}

Figure \ref{height} depicts the evolution of the height of the barrier 
$\Delta G_{max}$ as a function of misfit (in expanded overlayers, this 
is the highest value reached
before the collapse of the energy). The figures at each point show the number
of atoms in the cluster at the maximum of the transformation curve. As seen,
in the case of compressed overlayers, $\Delta G_{max}$ decreases steeply with
increasing misfit in a way similar to the decrease of the work required for 
nucleus formation
with increasing supersaturation in the classical theory of
nucleation.\cite{CGB} Assuming a dependence of the kind $\Delta G =
K\varepsilon ^{-n}$ where $K$ is a constant proportional to the Young modulus
(or the force constant $\gamma $) and $\varepsilon $ is the lattice misfit, we
found $n = 4.29$ for $\mu = 12, \nu = 6$, and $n = 4.75$ for $\mu = 36, \nu =
18$. It is worth noting that assuming 3D nucleation on top of the wetting
layer, Grabow and Gilmer predicted a value $n = 4$ for small misfits (large
nuclei) assuming that $\Delta G_{max}$ is inversely proportional to the 
square of the supersaturation, which in turn is proportional to the 
square of the lattice
misfit.\cite{GG} Note that the same exponent of four was obtained also by 
Tersoff and LeGoues.\cite{Jerr} Obviously, in our case the exponent $n$ is a
complicated function of the force constant in the interatomic bonds but the
value of the exponent is of the same order.

\begin{figure}[htb]
\includegraphics*[width=8.5cm]{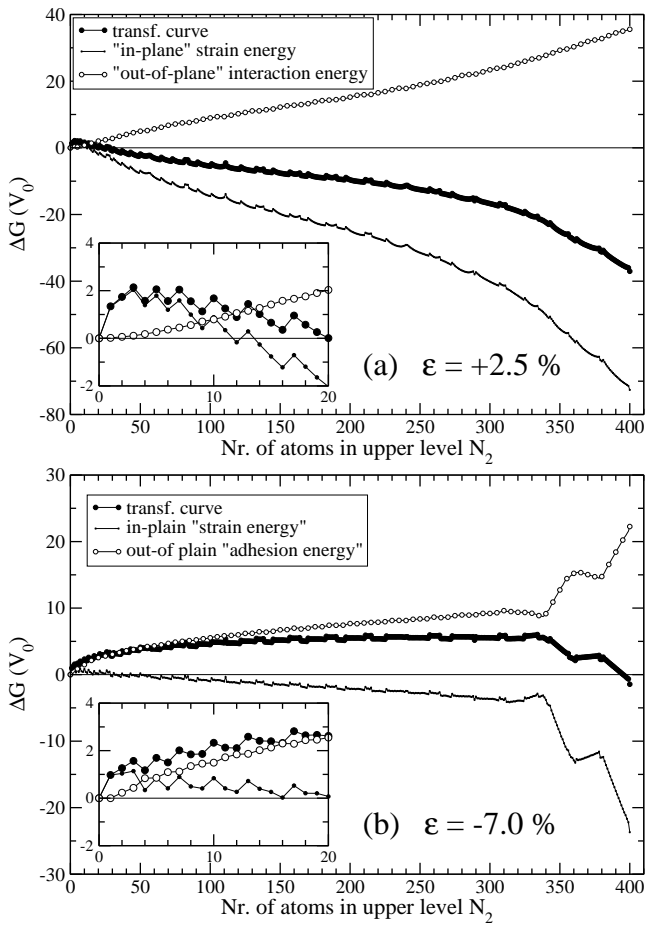}
\caption{\label{strains} Variation of the in-plane strain energy and
out\--of\--plane interaction energies during the mono-bilayer
trans\-for\-ma\-tion process in (a) compressed ($\varepsilon = +2.5~\%$) and
(b) expanded ($\varepsilon = -7.0~\%$) overlayers with $\mu=2\nu=36$.
The initial sizes of the islands were 29$\times$29=841 atoms. The inserts
show the curves at the beginning of the transformation with enlarged scales.}
\end{figure}

The misfit behavior of the critical nucleus size $n_{max}$ is also similar
to that found in the theory of nucleation. The nucleus size decreases with
increasing misfit, reaching eventually only one atom just as the nucleus size
on the su\-per\-sa\-tu\-ra\-tion in nucleation theory.\cite{CGB} It is
interesting to note that the critical nuclei contain always a number 
of atoms which {\em exceeds} by one atom the size of a compact cluster, 
and is {\em not} one atom short of being a compact cluster, as might be 
expected from too-simplistic energetic considerations on the basis of 
bond-counting arguments. The reason is that the highest departure from 
compactness
giving the maximum of the $\Delta G$ curve is achieved when the additional 
atom creates two kink positions for the attachment of the next row of atoms. 

On the contrary, in expanded overlayers, the number of atoms at which 
the transformation curve reaches its highest value does not depend 
on the misfit (Fig.~\ref{height}), de\-mon\-stra\-ting the 
non-nucleation behavior of the process.
This number is roughly equal to the number of atoms which completes the
upper level minus the number of atoms required to build the last four edge
rows of the upper level in order to produce four facets. A special feature of
the above results is that the bar\-ri\-ers for expanded overlayers 
are in general larger than those for com\-pres\-sed overlayers. 
Having in mind that the typical
time needed for the trans\-for\-mation to occur is inversely proportional to
$\exp(- \Delta G_{max}/kT)$, we have to expect much longer times
for the occurrence of the 2D-3D transformation in expanded overlayers as 
compared with compressed ones par\-ti\-cularly at larger values of the
misfit. Thus limitations of kinetic origin (astronomically long times for
second layer nucleation) are expected in expanded overlayers and at small
misfits in compressed overlayers.

Figure~\ref{n12dg} compares the misfit dependences of the critical island 
size $N_{12}$, the critical 2nd-layer nucleus size $n_{max}$ and the height
of the nucleation barrier $\Delta G_{max}$ for compressed overlayers. 
The tree curves display a similar
behavior increasing steeply with decreasing misfit, with $N_{12}$ showing 
a critical behavior at the critical misfit $\varepsilon _{12}$. 
As will be discussed below, the 2D-3D transformation will be inhibited 
for kinetic reasons at values of the misfit not sufficiently larger than
$\varepsilon _{12}$.

\section{Discussion}

In compressed overlayers the atoms interact with their in-plane neighbours
through the steeper repulsive branch of the interatomic potential. 
This means that compressed overlayers are effectively ``stiffer'' than 
ex\-pand\-ed overlayers. Then compressed pseu\-do\-mor\-phic overlayers 
will contain a larger amount of elastic energy than ex\-pand\-ed 
pseudomorphic ones of the same thickness and bonding strength. Another
consequence is that the accumulation of strain energy with thickness in
compressed overlayers will be steeper than in expanded overlayers. Or, the
strain energy of a bilayer will differ considerably from that of
a monolayer from both sides of the critical size $N_{12}$ compared with
expanded islands as seen in Fig. \ref{e12}. This is the reason why expanded
overlayers require greater force constants in order for the
monolayer islands to become unstable against bilayer islands. This is also
the reason why the critical sizes $N_{12}$ are larger in expanded than in
compressed overlayers of the same force constant as seen in Fig.~\ref{gamma}.

Figure~\ref{DeltaE} illustrates the central result of our study. It shows the
energy of transformation of compressed and expanded monolayer islands. 
In compressed overlayers the overall fall of the energy begins when a small 
cluster
(three atoms in this particular case) is formed in the second level. This is
a typical phase transition of first order - nucleation followed by
irreversible growth.\cite{Stmar} In expanded overlayers the energy increases
up to nearly the end of the transformation and then abruptly collapses. 
This collapse of the energy is connected with the transfer of the remaining 
last edge rows of atoms which leads to the coalescence of the lower and upper
steps to produce four side facets. 
In the particular transfer process considered in the calculation, there are
two regions, one before the beginning of the collapse, the second between
the two sub-collapses, where the total energy rises
slightly; this is due to the energetic cost of repulsion between steps that
are very close, separated only by a single atomic row.
The reason for this ``non-nucleation" 
behavior is the effectively weaker expanded bonds which results in relatively
close energies of the monolayer and bilayer islands. With increasing size of
the second level cluster, the misfit strain is not as effectively relaxed as 
in the case of compressed islands and the collapse of the energy is due to 
the replacements of isolated repulsing steps by low-energy facets.

The different transformation behavior can be un\-der\-stood on the basis of
the above considerations accounting in addition for the finite size of
the islands. Dur\-ing the transformation of the monolayer islands we should
ex\-pect a relaxation of the in-plane strain, which is the driving force 
for the transformation and an increase of the total energy of interaction 
between the lower level of the island and the substrate and between both 
levels of the island.\cite{Kor,Prieto}

The physics behind the process of mono-bilayer trans\-for\-mation is the same as
behind the formation of 3D islands on the wetting layer.~\cite{Jerr,Duport} 
Formation and growth of new steps and relaxation of the strain stored in the 
monolayer island compete in the process. In addition, the steps in the first 
and second levels repel each other and the associated repulsion energy 
increases towards the completion of the transformation to disappear 
when the steps coalesce to give rise to microfacets.

Epitaxial strain is relaxed at the islands edges.\cite{Vill} Edge atoms are
displaced from the bottoms of their respective potential troughs giving rise
to relaxation of the bonds parallel to the surface (in-plane stress
relaxation). The dis\-placed atoms loose contact with the substrate atoms,
which leads to an increase of the out-of-plane energy of interaction with the
underlying substrate.\cite{Rolf,JH}
During the transformation, new edges on top of the initial monolayer island
are formed and the total length of the edges increases as $\Delta
L(N_2) = 4(\sqrt{N_0 - N_2}+\sqrt{N_2}-\sqrt{N_0})$ where $N_0$ and $N_2$ 
are the number of atoms in the initial monolayer island and the current 
number of transferred atoms in the second level, respectively. Note that the 
energy of repulsion between the edges bounding the lower and upper islands is
implicitly accounted for by both the in-plane and out-of-plane energies. Then a
larger number of atoms are displaced from the bottom of their respective 
potential troughs during the transformation process and the total 
in-plane strain-relaxation energy decreases.
Simultaneously, the out-of-plane interaction energy increases. Owing to the
weaker attractive forces in expanded over\-lay\-ers, only a small number
of bonds close to the edges are relaxed.~\cite{Prieto} 
Most of the bonds at the center of
the islands are strained to fit the underlying wetting layer. As a result, the
average relaxation in expanded islands is smaller than in compressed
islands, where even bonds at the center of medium-sized islands are 
partly relaxed.
In compressed overlayers, the decrease of the in-plane strain energy rapidly
overcompensates the increase of the out-of-plane interaction energy which
results
in a nucleation-like transformation curve (Fig. \ref{strains}). In expanded
overlayers, the absolute value of the total in-plane strain energy is
smaller than the out-of-plane interaction energy with the exception of the
final stage when the monolayer-high steps disappear to produce facets of 
small surface energy.

The typical time required for the appearance of a second-layer nucleus is 
inversely proportional to the nucleation frequency
$\omega = S_{12}K\exp(- \Delta G_{max}/kT)$, where $S_{12} =
a^2N_{12}$ is the area of the critical monolayer island  and $K$ is the
pre-exponential of the nucleation rate.
As seen in Fig. \ref{height}, in the case of $\mu = 2\nu = 12$, the barrier 
height increases ap\-pro\-xi\-mately 5~times in an interval of $\varepsilon$ 
of 2.5~\% whereas the number of atoms in the critical nucleus $n_{max}$ 
increases nearly 70~times. 
For a greater force constant ($\mu = 2\nu = 36$), the increase of
both $\Delta G_{max}$ and $n_{max}$ is even larger: 20 and 110 times,
respectively, in a smaller misfit interval of about 1.5~\%. 
The energy to break a first-neighbor bond, $V_0$, for most semiconductor
materials is of the order of 2 to 2.5 eV (the enthalpy of evaporation is of
the order of 4 to 5 eV). Assuming $N_{12}$ is of the order of 100 -
120 atoms we could expect a mono-bilayer transformation to take place at
misfits for which $\Delta G_{max}/kT < 15 - 20$ ($n_{max} \le 3$). The reason
is that the pre-exponential $K$ in 2D nucleation rate from vapor is usually
of the order of $10^{20}$ cm$^{-2}$s$^{-1}$.\cite{CGB} 
Otherwise, due to the exponential dependence, times of the order of 
centuries would be required for second-layer nucleation.\cite{Dash}
Thus, although in compressed overlayers second-layer nucleation can be
expected for thermodynamic reasons at misfits above
$\varepsilon _{12}$, a real 2D-3D transition can only take place at even
larger misfits or higher temperatures for kinetic reasons. As the height of
the transformation barriers in expanded overlayers is always greater than
several times $V_0$, the mono-bilayer transformation should be strongly 
inhibited for kinetic reasons.

We conclude that the case of a layer-by-layer me\-cha\-nism for the 2D-3D
trans\-for\-mation is expected only in compressed overlayers at
mis\-fits suf\-ficiently larger than $\varepsilon _{12}$. The reason is that
the me\-cha\-nism of the mono-bilayer transformation is nuc\-le\-a\-tion-like
due to the interplay of relaxation of the in-plane strain, which is
proportional to the total edge length and the increase of the total edge
energy and repulsion between the edges. 
The transformation curve in expanded overlayers shows a
``non-nucleation" behavior cha\-racterized by an overall increase of the 
energy up to the stage when the single steps coalesce to produce low-energy
facets. The latter is accompanied by a col\-lapse of the energy. The maximum
energy is large and 2D-3D transformation of expanded overlayers is not
expected for kinetic reasons even for materials with strong interatomic bonds.
Softer materials are expected to grow with a planar morphology until misfit
dislocations are introduced, or to transform into 3D islands by a different
mechanism.

\acknowledgments

J.E.P. gratefully acknowledges financiation by the programme 
``Ram\'on y Cajal'' of the Spanish Ministerio de Educaci\'on y Ciencia.

\end{document}